\documentclass[aps,prl,secnumarabic,
groupedaddress,showpacs,amsmath,amssymb]{revtex4}
\usepackage[T2A]{fontenc}
\usepackage[cp1251]{inputenc}
\usepackage[english,russian]{babel}

\usepackage{bm}
\begin{document}
\title{INSTABILITIES IN ONE-DIMENSIONAL TWO-COMPONENT ATOMIC FERMI GAS WITH TWO-LEVEL IMPURITIES}
\author{Л.~А.~Манакова}
\email[]{manakova@kurm.polyn.kiae.su} \affiliation{ РНЦ
"Курчатовский Институт", Пл.Курчатова\ 1, 123182 Москва, Россия}

\begin{abstract}
In the present paper one-dimensional two-component atomic Fermi
gas is considered in long-wave limit as a Luttinger liquid. The
mechanisms leading to instability of the non-Fermi-liquid state of
a Luttinger liquid with two-level impurities are proposed. Since
exchange scattering in 1D systems is two-channel scattering in a
certain range of parameters, several types of non-Fermi-liquid
excitations with different quantum numbers exist in the vicinity
of the Fermi level. These excitations include, first, charge
density fluctuations in the Luttinger liquid and, second,
many-particle excitations due to two-channel exchange interaction,
which are associated with band-type as well as impurity fermion
states. It is shown that mutual scattering of many-particle
excitations of various types leads to the emergence of an
additional Fermi-liquid singularity in the vicinity of the Fermi
level. The conditions under which the Fermi-liquid state with a
new energy scale (which is much smaller than the Kondo
temperature) is the ground state of the system are formulated.

\end{abstract}

\maketitle

\section  {I. INTRODUCTION}

The interest in low-dimensional, in particular, one-dimensional (1D), systems has been
revived in recent years in connection with the obtaining of degenerate quantum gases
in quasi-one-dimensional magnetic and optical traps as well as in 1D optical lattices.
The previous peak of interest in 1D systems about ten years ago (see references sited
in [1, 2] and in this work) was due to the development of so-called quantum wires,
viz., 1D electronic systems in inversion layers of GaAs [3]. Degenerate atomic Fermi
gases have already been obtained in traps [4-7]. It should be noted that Fermi gases
with practically any number of particles and any intensity of the interaction between
them can be produced in traps. This can be done by using the Feschbach resonance [7].
Thus, we can expect that various remarkable properties predicted theoretically for 1D
systems both in the framework of the low-energy Luttinger model [8] as well as using
exactly solvable models (see, for example, [9]) can be experimentally observed in such
systems with a quasi-one-dimensional geometry. The properties of 1D metals in a fairly
wide range of parameters can be described using the Luttinger model (see, for example,
recent reviews [1, 2] and the literature cited therein). Accordingly, such materials
are referred to as Luttinger liquids (in contrast to the Fermi liquid in the 3D case).
In such systems, even a weak interaction leads to a qualitative rearrangement of the
excitation spectrum at low energies. Namely, 1D metals have no well-defined
one-particle excitations. The only stable excitations in the vicinity of the Fermi
level are collective charge and spin density fluctuations (acoustic modes). These
excitations are dynamically independent that corresponds to complete separation of the
spin and charge degrees of freedom. The interactions also lead to a power decay of all
correlation functions over large distances and times. The response to local
perturbations is an important problem for real systems. In this case, the behavior of
1D systems also differs qualitatively from the situation in 3D metals. It was shown in
the famous publication by Kane and Fisher [10] that the potential scattering of right
fermions by left ones (so-called backward scattering) in a Luttinger liquid with a
repulsive interaction leads to complete reflection of excitations from the potential
at low energies. The X-ray response in a Luttinger liquid has been studied extensively
[11-13] (including the situation with backward scattering [14]). The exchange
interaction with a spin impurity is one of the central problems for strongly
correlated systems [15]. It should be noted that if the number of electron channels
participating in the exchange interaction exceeds double the impurity spin, the system
has a non-Fermi-liquid (NFL) fixed point, exhibiting an anomalous behavior of heat
capacity and susceptibility [16, 17]. The Kondo effect in Luttinger liquids was
studied in [18-20]. It was shown that, as in the 3D case, the problem could be
renormalized to the strong-coupling limit. However, two distinguishing features are
observed. First, the Kondo effect in a Luttinger liquid exists both for the
antiferromagnetic and ferromagnetic interactions. Second, the system has three fixed
points, two of which correspond to the one-channel Kondo behavior and one exhibits the
two-channel (i.e., non-Fermi-liquid) behavior (we consider the impurity spin 1/2). The
conditions for a stable two-channel Kondo behavior relative to the exchange backward
scattering were obtained in [20].

It is important for subsequent analysis to emphasize that, in the absence of
interaction between fermions, the NFL state associated with a multichannel
(two-channel) exchange interaction is unstable to any mechanism removing the
degeneracy of the channels participating in exchange scattering. In particular, the
instability of the non-Fermi-liquid (NFL) state in the two-channel Kondo model upon
the introduction of anisotropy of exchange constants in different channels was
considered in [21]. The instability of the same state to potential scattering of
many-particle excitations with different quantum numbers in quantum-dimensional
structures and in metals containing impurities of d- or f-elements was demonstrated in
[22, 23].

In this study, the mechanisms leading to instability of the NFL state of a Luttinger
liquid with two-level (pseudospin) impurities are proposed. Since exchange scattering
in 1D systems is of the two-channel type in a certain range of parameters, several
types of NFL excitations exist in the vicinity of the Fermi level: density
fluctuations of the Luttinger liquid in the charge channel and many-particle
excitations generated by the two-channel exchange interaction in the pseudospin
channel. It will be shown below that allowance for resonant scattering of 1D fermions
(along with their potential backward scattering) from many-particle excitations
generated by the two-channel exchange interaction leads in our case to the emergence
of additional narrow Fermi-liquid resonances in the vicinity of the Fermi level (even
for very weak backward scattering) and to the instability of the NFL state in this
sense. A transition from the NFL to the FL state is accompanied by the emergence of a
new small energy scale and, as a consequence, by an anomalous increase in the density
of states at low energies.

\section{II. Impurity-free 1D atomic Fermi
gas as a Luttinger liquid}

{\bf 1}. Fermi gases in a trap are formed by fermion atoms of mass
$m$ with two intrinsic states [4-7]. The number of atoms in each
state is the same. If the atoms are cooled to a temperature below
the Fermi temperature $T_F$, they form a degenerate Fermi gas. The
system can be treated as effectively one-dimensional if the
characteristic energy of longitudinal motion is much smaller than
the characteristic separation $\omega_{\bot}$ between transverse
quantization levels. In other words, the condition
$\varepsilon_F\ll \omega_{\bot}$ ($\hbar=1$) must be satisfied. At
low temperatures, only s-type collisions are possible between
Fermi atoms with different spins. For this reason, the
interactions are characterized by a single parameter, scattering
length $a\ll l_{\bot}$, where $l_{\bot}=(1/m\omega_{\bot})^{1/2}$.
The effective 1D interaction can be represented in the form of
short-range potential with a characteristic value of  $g=2\pi
a/ml^2_{\bot}$ [4, 5]. Taking these limitations into account, we
write the Hamiltonian
\begin{equation} \label{eq:1}
H=\sum\limits_{\sigma}\int dx
\Psi^+_{\sigma}(x)\Big[\frac{-\hbar^2}{2m}\cdot
\partial_x^2+V_{ext}(x)\Big] \Psi_{\sigma}(x)+g\sum\limits_{\sigma\neq \sigma'}\int dx
\Psi^+_{\sigma}(x)\Psi^+_{\sigma'}(x)\Psi_{\sigma}(x)\Psi_{\sigma'}(x)
\end{equation}
In a box of length $\mathcal{L}$ with periodic boundary conditions
$\Psi_{\alpha \sigma}(x)=(1/\sqrt{L})\sum_k e^{ikx}a_{k\sigma}$
(it should be recalled here that the relations $N\rightarrow
\infty,\;\;\;L\rightarrow \infty,\;\;\; (2\pi N/L)=k_F$ hold in
the continuous limit, $N$ being the number of fermions and $k_F$
the Fermi momentum), we have
\begin{equation} \label{eq:3}
H=\sum\limits_{k\sigma}\varepsilon_k
a^+_{k\sigma}a_{k\sigma}+\frac{g}{\mathcal{L}}\sum\limits_{k_1,k_2,q}a^+_{k_1,\uparrow}a_{k_1-q,\uparrow}
a^+_{k_2,\downarrow}a_{k_2+q,\downarrow}.
\end{equation}
In a 1D system the Fermi "surface" is composed by two points, $\pm k_{F}$. In the
non-interacting case one has $k_{F}=\pi \hbar n/2$,
$n=(N_{\uparrow}+N_{\downarrow})/\mathcal{L}$. Then, let us introduce the left(L)-
(around $-k_{F}$) and right(R)- (around $+k_F$) moving fermion operators
$a_{\alpha,\sigma}(k)$, $\alpha=\pm\equiv L,R$ and the corresponding densities
$\rho_{\alpha,\sigma}(q)=\sum\limits_k a^+_{\alpha,k+q,\sigma}a_{\alpha,k,\sigma}$. If
we suppose, that {\bf (1)} all the states up to the Fermi energy are occupied and {\bf
(2)} free dispersion relation is linear, $\varepsilon_{k\alpha}=\alpha \hbar v_F
k,\;\;v_F=\pi \hbar n/2m$, we obtain
$$[\rho_{\alpha\sigma}(-q),\rho_{\alpha'\sigma'}(q')]=\delta_{\sigma,\sigma'}\delta_{\alpha,\alpha'}\delta_{q,q'}
\frac{q\alpha\mathcal{L}}{2\pi};\;\;\;[H_0,\rho_{\alpha}(q)]=q\alpha\rho_{\alpha}.$$
With taking these relations into account, the kinetic energy takes the form
$$H_0=\frac{2\pi
v_F}{\mathcal{L}}\sum\limits_{\alpha,q>0,\sigma}\rho_{\alpha,\sigma}(q)\rho_{\alpha,\sigma}(-q)=
\pi v_F\sum\limits_{\alpha,\sigma}\int dx
\rho_{\alpha,\sigma}^2(x);\;\;\rho_{\alpha,\sigma}(x)=\Psi^+_{\alpha,\sigma}(x)\Psi_{\alpha,\sigma}(x).$$
In turn, the interaction Hamiltonian can be written as
$$H_{int}=\sum\limits_{\alpha\neq\alpha', \sigma\sigma'}\Big[\int dx
\Big(g_2^{\sigma\sigma'}\rho_{\alpha,\sigma}(x)\rho_{\alpha'\sigma'}(x)+g_4^{\sigma\sigma'}
\rho_{\alpha,\sigma}(x)\rho_{\alpha\sigma'}(x)\Big)+ g_1^{\sigma\sigma'}\int dx
\Psi^+_{\alpha,\sigma}(x)\Psi^+_{\alpha',\sigma}(x)\Psi_{\alpha,-\sigma}(x)
\Psi_{\alpha',-\sigma}(x)\Big].$$

For $g_1\geq 0$ one renormalizes to the fixed line $g_1^*=0,\;\;g_2^*=g_2-g_1/2$ and
the fixed point Hamiltonian is a Luttinger model [1,2]. In this case $g_1$-interaction
is irrelevant. For spin independent interactions one has $g_{2c}^*\neq
0,\;\;g_{2s}^*=0$ and $K^*_s=1$. These relations are generated by the equations
$g_{2,c}^*=g^{*\sigma\sigma}_2+g^{*\sigma-\sigma}_c;\;\;\;
g_{2,s}^*=g^{*\sigma\sigma}_2-g^{*\sigma-\sigma}_2$. In what follows, the subscript
$*$ is omitted.

The Hamiltonian $H_0+H^{\mathcal{L}}_{int}$ with $H^{\mathcal{L}}_{int}\equiv
H_{int}(g_2,g_4)$ is equivalent to that of Luttinger model. To obtain the standard
form of the Luttinger Hamiltonian, the densities $\rho_{\alpha,\sigma}(x)$ are written
by means of canonically cojugate boson fields $\phi_{\nu}$ and $\theta_{\nu}$
\cite{S}, \cite{SCP}, \cite{Hald}: $\rho_{\alpha\sigma}=(1/\sqrt{8\pi})\Big(\partial_x
\phi_c-\alpha\Pi_c+\sigma(\partial_x \phi_s-\alpha\Pi_s)\Big)$. Here $ \Pi_{\nu}\equiv
\partial_x \theta_{\nu}$, $[\phi_{\nu}(x),\phi_{\nu'}(x')]=0$,
$[\Pi_{\nu}(x),\Pi_{\nu'}(x')]=0$,
$[\phi_{\nu}(x),\Pi_{\nu'}(x')]=i\delta_{\nu,\nu'}\delta(x-x')$, $\nu=c,s$. The fields
$\phi_{c,s}$ are determined by charge and spin fluctuations respectively:
$\rho_c=\partial_x \phi_c/\sqrt{\pi}$, $\rho_s=\partial_x \phi_s/\sqrt{\pi}$.

It should be emphasized that {\bf in an atomic Fermi gas, "charge" fluctuations
correspond to fluctuations of the average density of the gas, while "spin"
fluctuations describe the fluctuations of the relative density at two levels
corresponding to intrinsic states of atoms.} In what follows we use the standard
terminology of a Luttinger model.

The Hamiltonian $H_0+H^{\mathcal{L}}_{int}$ takes the form
\begin{equation}
\begin{split}
&H_{\mathcal{L}}=H_c+H_s;\;\;\;H_{\nu}=\int dx \frac{u_{\nu}}{2}
\Big( K_{\nu}\Pi_{\nu}^2+\frac{1}{K_{\nu}}(\partial_x
\phi_{\nu})^2\Big);\;\;\;v_s=v_F;\;\;\;K_s=1\\
&\;\;\;\;\;\;\;\;\;\;\;\;\;u_{c}=\sqrt{\Big(v_F+\frac{g_{4}}{\pi}\Big)^2-\Big(\frac{g_{2}}{\pi}\Big)^2};\;\;\;
K_{c}=\sqrt{\frac{\pi v_F+g_{4}-2g_{2}}{\pi v_F+g_{4}+2g_{2}}}.
\end{split}
\label{eq:8}
\end{equation}

In the Luttinger model fermion fields $\Psi_{\alpha\sigma}(x)$ are defined as (see,
for example, \cite{SCP}, as well as \cite{Hald})
\begin{equation}
\Psi_{\alpha\sigma}(x)=\frac{\eta_{\alpha\sigma}}{\sqrt{2\pi a
}}\exp\Big[-i\sqrt{4\pi}\Phi_{\alpha\sigma}(x)\Big];\;\;\;
\Phi_{\alpha\sigma}(x)=\frac{1}{\sqrt{2}}\Big[\alpha\Big(\phi_c(x)+
\sigma\phi_s(x)\Big)+\Big(\theta_c(x)+\sigma\theta_s(x))\Big],
\label{eq:4}
\end{equation}
$\eta_{\alpha\sigma}$ are the so-called Klein factors.

For a pure Luttinger liquid, the Green function $\mathcal{G}_{\alpha\sigma}(x;t)$ was
obtained in \cite{LuP}. The spectral functions and, accordingly, the density of states
$\rho_{0\alpha\sigma}(\varepsilon)$ were calculated almost twenty years later in
\cite{MS}, \cite{SM}. In particular, the total spectral function (density of states)
for a pure Luttinger liquid in the long-wave limit is defined as
\begin{equation}
\rho_{0\alpha\sigma}(\varepsilon)=
\frac{A}{\Gamma(1+\eta_c+\eta_s)\varepsilon_0}\cdot\Big(\frac{|\varepsilon|}{\varepsilon_0}\Big)^{\eta_c+\eta_s};\;\;\;\;
\eta_{\nu}=\frac{(K_{\nu}-1)^2}{4K_{\nu}};\label{eq:28}
\end{equation}
Here, $A\sim 1$, $\varepsilon_0\sim\varepsilon_F$ is the cutoff energy in the
Luttinger model; $\eta_c,\eta_s$ are anomalous dimensions in the charge and pseudospin
channels; and parameters $K_c, K_s$ are defined in (\ref{eq:8}). The energy is
measured from the Fermi level.

The Hamiltonian (\ref{eq:8}) also describes long-wave excitations in one-dimensional
optical lattice with the number of fermions per site smaller than unity (metallic
state).

The possibility of describing an atomic impurity-free Fermi gas with the help of a
Luttinger model and the methods for experimental observation of the spin-charge
separation were also considered in \cite{RFZZ}.

\section {IMPURITY HAMILTONIAN AND SCATTERING PROBLEM}

1. Let us consider the situation when a 1D lattice with a number of fermion per site
smaller than unity (metallic state) contains localized fermions with two internal
degrees of freedom. The energy of a localized state is much lower than the Fermi
level. Localized fermions do not interact with one another. We assume, however, that
the wave functions of localized and band fermions may overlap; for this reason, a
localized state interacts with band fermions. Since a two-level atom can be described
by a pseudospin variable with two values of the z component, we will refer to this
situation as a pseudospin impurity in a metallic 1D lattice. We assume that the
repulsion of fermions at an impurity site is so strong that only one fermion can
occupy this site. We will consider below a 1D system with periodic boundary
conditions. In the case of an atomic Fermi gas, the pseudospin degrees of freedom in
the optical lattice correspond to two intrinsic atomic levels of the hyperfine
structure. The Hamiltonian of such a system will be written in the form
\begin{equation} \label{eq:10a}
\begin{split}
&H=\mathcal{H}+H_{sc};\;\;\;\mathcal{H}=
H_{\mathcal{L}}+H_{0d}+H_{int};\;\;\;
H_{0d}=\sum_{\sigma}E_{d}n_{d\sigma};\;\;\;H_{sc}=H_{bs}+H_h\\
&\;\;H_{bs}=\sum _{kk'}\sum_{\alpha\neq\alpha',\sigma}
\left(T^{\alpha \alpha'}_{kk'}
a^+_{k\alpha\sigma}a_{k'\alpha'\sigma} + H.c.\right);\;\;\;
H_{h}=\sum_{k\alpha\sigma}\left(V^{\alpha}_{kd}
a^+_{k\alpha\sigma}d_{\sigma} + H.c.\right),
\end{split}
\end{equation}
where $H_{\mathcal{L}}$ is the Hamiltonian of 1D fermions, which coincides with the
Hamiltonian in the Luttinger model in the continuous limit, $H_{0d}$ is the
Hamiltonian of a localized level with energy $E_d$,
$n_{d\sigma}=d^+_{\sigma}d_{\sigma}$, $H_{h}$ is the hybridization between band and
localized fermions; $H_{bs}$ describes the potential scattering of band fermions with
different quantum numbers (in other words, backward scattering). The term $H_{int}$
from Eq. (\ref{eq:10a}) corresponds to the repulsion between fermions at a deep level,
$$H_{int}=\frac{1}{2}U\sum_{\sigma}n_{d\sigma}n_{d-\sigma}.$$ The interaction between
band fermions and a localized state is generated by $H_{int}$ with taking
hybridization into account.

The results obtained in what follows are applicable to an atomic gas if, in addition
to the relations given above, the inequality $U<\omega_{\bot}$ holds.

{\bf 2.} Excitations in a system with Hamiltonian $H=\mathcal{H} +
H_{sc}$ are completely determined by the Green function of complex
argument z, $\mathbf{G}_{d}(z)=<d|(z-\hat{H})^{-1}|d>$, which can
be evaluated using the equations of motion. The system of
equations of motion for $\mathbf{G}_{d}(z)$ has the form :
\begin{equation}
\begin{split}
&\mathcal{G}^{(0)-1}_{d}(z)\mathbf{G}_d(z)-\sum_{p\alpha=L,R}
V^{\alpha}_{dp}\mathbf{G}_{d\alpha}(p;z)=\hat{1};\\
&\mathbf{G}^{-1}_L(p;z)\mathbf{G}_{dL}(p;z)-\sum_{p'}
T^{LR}_{pp'}\mathbf{G}_{dR}(p';z)-V^{L}_{pd}\mathbf{G}_{d}(z)=0;\\
&\mathbf{G}^{-1}_R(p;z)\mathbf{G}_{dR}(p;z)-\sum_{p'}
T^{RL}_{pp'}\mathbf{G}_{dL}(p';z)-V^{R}_{pd}\mathbf{G}_{d}(z)=0;
\end{split}
\label{eq:18}
\end{equation}
Here $\mathcal{G}^{(0)}_{d}(z)=<d|[z-(H_{0d}+H_{int})]^{-1}|d>$,
$\mathbf{G}_{\alpha}(p;z)=<a_{p\alpha}|(z-\hat{\mathcal{H}})^{-1}|a_{p\alpha}>$,
$\mathbf{G}_{d\alpha}(k;z)=<d|(z-\hat{\mathcal{H}})^{-1}|a_{k\alpha}>$ are the Green
functions taking into account all interactions, but disregarding scattering. In
expressions (\ref{eq:18}) and below, the spin indices are temporarily omitted. In the
case of separable matrix elements $T^{\alpha\alpha'}_{kk'}$ and/or those exhibiting a
weak energy dependence the solution for $\mathbf{G}_{d}(z)$ has the form
\begin{equation}
\begin{split}
&\mathbf{G}_{d}(z)=\frac{\mathcal{G}_{d}(z)}{1-\Sigma_{sc}(z)\mathcal{G}_{d}(z)};\;\;\;\;
\mathcal{G}^{-1}_{d}(z)=\mathcal{G}^{(0)-1}_{d}(z)-\sum_{\alpha}\Sigma^{VV}_{\alpha}(z),\\
\Sigma_{sc}(z)=&\frac{\Sigma^{VT}_L(z)\Big[\Sigma^{TV}_R(z)+\Sigma^{TT}_R(z)\cdot\Sigma^{TV}_L(z)\Big]+
\Sigma^{VT}_R(z)\Big[\Sigma^{TV}_L(z)+\Sigma^{TT}_L(z)\cdot\Sigma^{TV}_R(z)\Big]}{1-
\Sigma^{TT}_L(z)\Sigma^{TT}_R(z)};\\
&\Sigma^{ab}_{\alpha}(z)=\sum_p
W^{ab}_{\alpha}(p)\mathbf{G}_{\alpha}(p;z)=
W^{ab}_{\alpha}(p_F)\int\limits d\varepsilon
\frac{\rho_{\alpha}(\varepsilon)} {z-\varepsilon}\equiv
W^{ab}_{\alpha}(p_F)\Sigma_{\alpha}(z),
\end{split}
\label{eq:19}
\end{equation}
here $\mathcal{G}_{d}(z)$ is the Green function of impurity degrees of freedom
disregarding scattering processes associated with $H_{bs}$. However, by definition,
this function includes all interactions between band and localized fermions, which are
generated by the repulsion $H_{int}$ of fermions at an impurity site. The self-energy
functions $\Sigma^{ab}_{\alpha}(z)$ are written in the form of spectral expansions of
many-particle Green functions for 1D fermions; $W^{ab}_{\alpha}$ are the products of
matrix elements, which are defined by the superscripts of the self-energy functions;
and $\rho_{\alpha}(\varepsilon)$ is the density of states corresponding to the
many-particle excitation spectrum taking into account the interactions. This spectrum
is determined by the Green function $\mathbf{G}_{\alpha}(p;z)$. The complete Green
function in relations (\ref{eq:19}) has singularities of two types. Functions
$\mathcal{G}_{d}(z)$ contains singularities generated by the interaction between band
fermions and a localized state. The denominators in relations (\ref{eq:19}) appear due
to scattering of various types of many-particle excitations from one another. It
follows from the expression for $\Sigma_{sc}(z)$ that the system experiences, first,
potential scattering of right fermions from left ones (the term containing
$\Sigma^{TT}_L(z)\Sigma^{TT}_R(z)$, but taking into account the modification of the
density of states due to the exchange interaction and, second, the resonance
scattering of 1D fermions from many-particle excitations, which are described by the
terms with the Green function $\mathcal{G}_{d}(z)$. It will be shown below that in
both cases scattering may lead to the emergence of additional singularities (namely,
simple poles) in the Green function in the vicinity of the Fermi level. The poles
correspond to Fermi-liquid excitations. The position of poles
$z_r=\varepsilon_r+i\gamma_r$ is determined from the solution of the equation

\begin{equation}
1-\Sigma_{sc}(z_r)\cdot
\mathcal{G}_{d}(z_r)=0.
\label{eq:20}
\end{equation}

Thus, to solve the scattering problem, we must know the many-particle low-energy
density of states for 1D fermions (disregarding scattering) as well as the Green
function $\mathcal{G}_{d}(z)$ for the resonance level.

{\bf 3.} In the present paper, assuming that the interactions of fermions with one
another and with the impurity are strong, we first determine low-energy many-particle
excitations, and then take into account the scattering of these excitations from one
another. In other words, first the problem with Hamiltonian $\mathcal{H}$ is solved
and $\mathcal{G}_{d}(z)$ is determined with the help of the methods used in the
problem with $U\rightarrow \infty$.

Because the energy $U$ is dominant, it is essential to properly treat correlation on
site, and the method of the equations of motion correctly accounts for these on-site
correlations \cite{H}. This method gives the following expression for the Green
function of band fermions
\begin{equation}
\mathbf{G}_{\alpha\alpha\sigma}(k,k';z)=\delta(k-k')\mathcal{G}_{\alpha\sigma}(k;z)
+\mathcal{G}_{\alpha\sigma}(k;z)\mathcal{T}_{\alpha\alpha}(k,k';z)\mathcal{G}_{\alpha\sigma}(k';z),
\label{eq:22}
\end{equation}
where $\mathcal{G}_{\alpha}(k;z)$ is the Green function of a pure Luttinger liquid,
and the scattering matrix has the form
$$\mathcal{T}_{\alpha\alpha}(k,k';z)=V^{*\alpha}_{kd}\mathcal{G}_{d}(z)V^{\alpha}_{k'd},\;\;\;
\mathcal{G}_{d}(z)=\mathcal{G}^{(0)-1}_{d}(z)-\sum_{\alpha}\Sigma^{VV}_{0\alpha}(z),$$

Supposing that in the limit of large values of U the main effect of this interaction
is the many-particle resonance at the Fermi level, we use the following relation for
low energies
\begin{equation} \label{eq:22a}
\mathcal{G}^{-1}_{d}(z)=\mathcal{G}^{(0)-1}_{d}(z)-\sum_{\alpha}\Sigma^{VV}_{0\alpha}(z),
\approx <d|[z-(H_{\mathcal{L}}+H_{0d}+H_{ex}+H^{(0)}_{bs})]|d>,
\end{equation}
here $\Sigma^{VV}_{0\alpha}(z)$, $H^{(0)}_{bs}$ are defined with
$\rho_{0\alpha\sigma}(\varepsilon)$ corresponding to a pure Luttinger liquid, $H_{ex}$
is the exchange scattering of the 1D fermions by two-level (psevdospin) impurity. In
accordance with the relation (\ref{eq:22a}), to calculate $\mathcal{G}_{d}(z)$ in what
follows, we represent the exchange interaction in the form of the model of a resonance
level.

By means of Eq. (\ref{eq:22}), the density of states is defined as
\begin{equation}
\rho_{\alpha\sigma}(\varepsilon)=\mp\pi^{-1} {\rm
Im}\Big[Sp\mathbf{G}_{\alpha\alpha\sigma}(k,k';\varepsilon)\Big]=\rho_{0\alpha\sigma}(\varepsilon)\mp
\pi^{-1} {\rm Im}\Big[Sp\mathcal{G}_{d}(\varepsilon)
\sum_k|V^{\alpha}_{kd}|^2\mathcal{G}^2_{\alpha\sigma}(k;\varepsilon)\Big]\equiv
\rho_{0\alpha\sigma}+\rho_{d\alpha\sigma} , \label{eq:23}
\end{equation}
where the minus and plus signs correspond to $\varepsilon>0$ and
$\varepsilon<0$, respectively. The relation (\ref{eq:22a}) implies
that only the density of states $\rho_{d\alpha\sigma}$ generated
by the exchange scattering enters the self-energy $\Sigma_{sc}(z)$
in Eq. (\ref{eq:19}).

\section {The resonance-level model for the exchange interaction}

In one-dimensional systems the exchange interaction takes the form
$H_{ex}=H^{(F)}_{ex}+H^{(B)}_{ex}$, where
\begin{equation}
H^{(F)}_{ex}=\sum_{\alpha}\sum_{i=x,y,z}\sum_{\sigma,\sigma'}J^i\Psi^+_{\alpha
\sigma}(0)\hat{\tau}^i_{\sigma,\sigma'}\Psi_{\alpha
\sigma'}(0){\bf S}^i;\;\;\;
H^{(B)}_{ex}=\sum_{\alpha\neq\alpha'}\sum_{i=x,y,z}\sum_{\sigma,\sigma'}J_B^i\Psi^+_{\alpha
\sigma}(0)\hat{\tau}^i_{\sigma,\sigma'}\Psi_{\alpha'
\sigma'}(0){\bf S}^i \label{eq:11}
\end{equation}
Here, ${\bf S}$ is the impurity pseudospin, and ${\tau}^i$ are the
Pauli matrices. Hamiltonian $H^{(B)}_{ex}$ corresponds to the
back-ward exchange scattering emerging in the presence of
anisotropy of scattering channels, $H^{(F)}_{ex}$ describes the
two-channel exchange scattering.

The interaction in the charge channel of a Luttinger liquid is characterized by
parameter $K_c$ (see expression (\ref{eq:9}) below). It was proved in [20] that, in
the case of repulsive interactions ($K_c < 1$), the small anisotropy ($J_B$) of the
exchange scattering channels in the vicinity of the fixed point of strong interaction
in $J$ is relevant only for $1/2 < K_c < 1$. In this case, the results obtained in
[19] are valid (it was found in [19] that the two-channel Kondo model is absolutely
unstable). However, for $K_c < 1/2$, the two-channel Kondo behavior is stable to the
exchange anisotropy of the channels. In this case, however, one more mechanism
violating the channel degeneracy (weak resonant and potential scattering of
many-particle excitations) in the two-channel exchange interaction was discarded. We
will consider this mechanism here. Using the fact that the exchange anisotropy of the
channels is irrelevant for $K_c < 1/2$, we consider the situation when the impurity
pseudospin has a symmetric coupling with adjoining lattice sites so that $J_B= 0$. It
will be shown below that Hamiltonian (\ref{eq:11}) in this case can be reduced to the
model of the resonance level (this model was constructed in \cite{EK} for the
two-channel exchange scattering in metals disregarding an interaction between band
electrons).

For 1D system, in order to represent the two-channel exchange interaction in the form
of the model of a resonance level, we introduce the boson representation (\ref{eq:4})
of fermion fields $\Psi_{\alpha\sigma}(x)$. Hamiltonians $H_{\nu}$ in Eq. (3) lead to
the canonical (for bosons) form $$H'_{\nu}=\frac{u_{\nu}}{2}\int dx
\Big(\Pi^{'2}_{\nu}+(\partial_x \phi_{\nu}')^{2}\Big)$$ with the help of the following
redefinition of boson fields: $\phi_{\nu}=\sqrt{K_{\nu}}\cdot
\phi'_{\nu};\;\;\;\Pi_{\nu}=(1/\sqrt{K_{\nu}})\cdot \Pi'_{\nu}$. In this case, phases
$\Phi_{\alpha s}$  in relation (\ref{eq:4}) are transformed to
\begin{equation}
\Phi'_{\alpha
\sigma}=\frac{1}{\sqrt{2}}\Big[\pm\Big(\phi'_c\sqrt{K_c}+
\sigma\phi'_s\sqrt{ K_s}\Big)+\Big(\frac{\theta'_c}{\sqrt{
K_c}}+\sigma\frac{\theta'_s}{\sqrt{K_s}}\Big)\Big].
\label{eq:10}
\end{equation}
Substituting the fields $\phi'_{\nu},\;\theta'_{\nu}$ into Eq. (\ref{eq:11}) and
taking into account relation (\ref{eq:10}), we obtain
\[
\begin{split}
&\sum_{\sigma,\sigma'}J_{\bot}\psi'^+_{\alpha
\sigma}(0)\hat{\tau}^i_{\sigma,\sigma'}\psi'_{\alpha \sigma'}(0)
S^+= \frac{J_{\bot}}{2\pi a}S^+\Big[\exp
[i\Big(\phi'_{s}\sqrt{4\pi K_{s}}-
\sqrt{\frac{4\pi}{K_{s}}}\cdot\theta'_{s}\Big)]+\\&+\exp[-i\Big(\phi'_{s}\sqrt{4\pi
K_{s}}
+\sqrt{\frac{4\pi}{K_{s}}}\cdot\theta'_{s}\Big)]\Big]_{x=0};
\;\;\; J_z S^z\sum_{\alpha}(\rho_{\alpha \uparrow}-\rho_{\alpha
\downarrow})=J_z S^z\sqrt{\frac{2}{\pi K_{s}}}(\partial_x
\theta'_{s})_{x=0}.
\end{split}
\]

Thus, in the general case, the exchange interaction in a Luttinger
liquid has the form
\begin{equation} \label{eq:12}
H_{ex}=\frac{J_{\bot}}{\pi a}\Big[S^+\cos\Big(\sqrt{4\pi
K_{s}}\cdot\phi'_{s}(0)\Big)e^{-i\sqrt{(4\pi/K_{s})}\cdot\theta'_{s}(0)}+H.c.\Big]+
J_z S^z\sqrt{\frac{2}{\pi K_{s}}}(\partial_x \theta'_{s})_{x=0}.
\end{equation}

It should be emphasized that the expression for the exchange interaction contains {\bf
only pseudospin fields}. To reduce expression (\ref{eq:12}) to the Hamiltonian in the
model of a resonant level, we carry out the following transformations.
\[
\begin{split}
&1.\;\;\text{We introduce fields }\;\;\Phi_{L,R}\;\;\text{instead
of}\;\;\phi'_{s},\;\;\theta'_{s}:\;\;
\phi'_{s}=\frac{\Phi_{L}+\Phi_{R}}{\sqrt{4\pi
K_{s}}};\;\;\;\theta'_{s}=(\Phi_{L}+\Phi_{R})\cdot\sqrt{\frac{K_{s}}{4\pi}}.\\
&2.\;\;\text{For convenience, we replace the right field
}\;\;\Phi_{R}\;\; \text{by the left field}\;\;
\Phi'_{L}:\;\;\Phi'_{L}(x)=-\Phi_{R}(-x).\\ &3.\;\;\text{We
introduce the symmetric and antisymmetric
fields}\;\;\Phi_{S,A}:\;\;
\Phi_{L}=\frac{\Phi_{S}+\Phi_{A}}{\sqrt{2}};\;\;\;\Phi'_{L}=\frac{\Phi_{S}-\Phi_{A}}{\sqrt{2}}.
\end{split}
\]
As a result of these transformations, Hamiltonian (\ref{eq:12}) assumes the form
\begin{equation} \label{eq:13}
H_{ex}\equiv H^{(s)}_{ex}=\frac{J_{\bot}}{2\pi a
}\Big[S^+\Big(e^{-2i(\Phi_S+\Phi_A)}+e^{-2i(\Phi_S-\Phi_A)}\Big)+H.c.\Big]+
\frac{J_z}{2\pi} S^z\cdot(\partial_x \Phi_S)_{x=0}.
\end{equation}
Field $\Phi_S$ from the transverse part of the interaction is eliminated by rotation
about the $S^z$ axis: $U=e^{2i\Phi_S S^z}$. For $g_1=0$ (or, which is the same, for
$K_s=1$), the kinetic energy in the pseudospin channel can be reduced to the form
corresponding to free boson fields so that
\begin{equation}
H_{s}=\frac{v_F}{\pi}\int
dx\Big(\Big[\nabla\Phi_S(x)\Big]^2+\Big[\nabla\Phi_A(x)\Big]^2\Big).
\label{eq:14}
\end{equation}
Since $UH_{s}U^{-1}=H_{s}-4(v_F/\pi)(\partial_x \Phi_S)S^z$, we
obtain, using the transformation $U(H_{s}+H^{(s)}_{ex})U^{-1}$ ,
the Hamiltonian  $H_s=H_{s}+H^{(s)}_{ex}$ in the pseudospin
channel in the model of a resonant level; here,
\begin{equation} \label{eq:15}
\begin{split}
&H^{(s)}_{ex}=\frac{J_{\bot}}{\sqrt{2\pi a }}\Big[\psi^+_A(0)+\psi_A(0)\Big](d^+-d))+
\frac{1}{2\pi}(J_z-8v_F) \cdot\psi^+_S(0)\psi_S(0)(d^+d-1/2)\equiv H_A+H_S;\\
&\psi^+_{S,A}(x)=\hat{\eta}\frac{e^{-2i\Phi_{S,A}(x)}}{\sqrt{2\pi
a}};\;\;\;S^+=d^+\hat{\eta};\;\;\;S^z=(d^+d-1/2);\;\;\;\hat{\eta}^2=1.
\end{split}
\end{equation}
It should be noted that phases $\Phi_{S,A}(x)$  describe the symmetric and
antisymmetric fluctuations of pseudospin density. The total Hamiltonian of a Luttinger
liquid with a two-level impurity has the form
\begin{equation}
H=H_c+H_{s}+H^{(s)}_{ex}+H_{sc}\equiv H_0+H^{(s)}_{ex}+H_{sc}.
\label{eq:16}
\end{equation}
If we disregard $H_{sc}$, Hamiltonian (\ref{eq:16}) gives
excitations of the Luttinger liquid with velocity $u_c$ in the
charge channel. In the pseudospin channel, excitations
$\psi^+_{S,A}$ are determined by the two-channel exchange
interaction, generating, among other things, a resonant
many-particle level for low energies. A remarcable feature of the
resonance-level model (\ref{eq:15}) is that the hybridization and
interaction are performed via different channels. This is its
essential difference from the resonance-level model for the
single-channel Kondo scattering, in which both of them are in the
same channel.

{\bf 2.} The model (\ref{eq:15}) can be renormalized to the limit of strong coupling
with the fixed point being on the line $\tilde{J}_z=(J_z-8v_F) = 0$ [21] (the
Emery-Kivelson line). On this line, the impurity degrees of freedom are hybridized
only with the field associated with antisymmetric pseudospin fluctuations. For $T =
0$, the Green function in the vicinity of the Fermi level has the form \cite{EK}
\begin{equation}
\mathcal{G}_{d}(\varepsilon)=\frac{1}{2}\left[\frac{\hat{\tau}_0-
\hat{\tau}_x}{\varepsilon+i\Gamma_{K} \text{sign} \varepsilon}+
\frac{\hat{\tau}_0+\hat{\tau}_x} {\varepsilon+i\delta \text{sign}
\varepsilon}\right].
\label{eq:25}
\end{equation}
In the strong coupling limit, $T_K\sim \Gamma_K$,  $T_K$ being the
Kondo temperature. The second term in relation (\ref{eq:25})
emerges due to the fact that half the impurity degrees of freedom
in the two-channel Kondo model are not hybridized with collective
variables of band fermions. Since Hamiltonian (\ref{eq:15}) does
not conserve the number of fermions, $\mathcal{G}_d (z)$ has
nonzero anomalous matrix elements $\sim <dd>$ and $\sim <d^+
d^+>$.

For the problem discussed in this paper, the form of $\mathcal{G}_d (z)$ beyond the
Emery-Kivelson line or, what is the same, for a finite interaction in the S-channel,
i.e., for $J_z-8v_F\neq 0$, is important. To obtain a solution for this case, let us
use the technique that was previously applied to the well-known problem of x-ray
absorption in metals \cite{SS}. First, the Hamiltonian $H_{0S}+H_S$, here
$H_{0S}=iv_F\int_{-\infty}^{+\infty} \psi^+_{S}(x)\partial_x\psi_{S}(x)$, of Eq.
(\ref{eq:15}) is diagonalized. To this end, we introduce boson operators
$b_{sk}=k^{-1/2}\rho_s(k)$, $b^+_{sk}=k^{-1/2} \rho_s(-k)$, where $\rho_s(k)$ are
density operators $$\rho_s(k)=\frac{1}{N^{1/2}}\sum_{q=0}^{k_D-k}
\psi^+_s(q)\psi_s(q+k):\;\; \rho_s(-k)=\frac{1}{N^{1/2}}\sum_{q=k}^{k_D}
\psi^+_s(q)\psi_s(q-k),\;\;k\ge 0,$$ $\psi_s(k)$ are Fourier components of fields
$\psi_s(x)$, and the cut-off takes place at $k_D\sim a^{-1}$. Using operators
$b_{sk}$, $b^+_{sk}$, we write the Hamiltonian $H_{0S}+H_S$ as

\begin{equation}
H_{0S}+H_S= v_F\sum_{k>0} kb^+_{sk}b_{sk}+
\lambda_z(d^+d-\frac{1}{2})\sum_{k>0}k^{1/2}(b^+_{sk}+b_{sk}), \label{eq:8.9}
\end{equation}

Here $\lambda_z\equiv (1/2\pi)(J_z-8\varepsilon_F)N^{-1/2}$. This Hamiltonian is
diagonalized to become $v_F\sum_{k>0} kb^+_{sk}b_{sk}$ by the canonical transformation

$$U=\exp\left(\lambda_z\rho_{0}(d^+d-\frac{1}{2}) \sum_{k>0}
k^{-1/2}(b_{sk}-b^+_{sk})\right),\;\;\rho_{0}\sim \varepsilon_F^{-1} .$$ In this
operation, the Hamiltonian $H_A$ is transformed to
\begin{equation}
\begin{split}
&\tilde{H}_{A}=\frac{J_{\bot}}{(2 a \pi)^{1/2}}[\psi^+_{A}(0)+\psi_{A}(0)]
(\tilde{d}^+-\tilde{d})+\Delta(\tilde{d}^+\tilde{d}-\frac{1}{2}),\\ &
\tilde{d^+}=Ud^+U^{-1}=\exp\left(\lambda_z\rho_{0a} \sum_{k>0}
k^{-1/2}(b_{sk}-b^+_{sk})\right)d^+\equiv U_0d^+,
\end{split}
\label{eq:8.10}
\end{equation}
$\Delta=-\varepsilon_J=\lambda_z^2\rho_{0}$ is the "polaron shift" due to the
screening interaction in the S-channel. With due account of Eq. (\ref{eq:8.10}), the
Green function $\mathcal{G}_d (t)$ of resonance level is

\begin{equation}
\mathcal{G}_d(t)=\mathcal{G}^0_d(t)\langle U^+_0(t)U_0(0)\rangle_D, \label{eq:8.11}
\end{equation}
$U_0(t)$ is derived from $U_0(0)$ using the substitution $b_{sk}\rightarrow
b_{sk}e^{i\varepsilon_kt}$. Here  $\langle...\rangle_D$ denotes averaging over the
states of the diagonalized Hamiltonian $H_{0S}+H_S$, $\mathcal{G}^0_d (t)$ is the
Fourier transform of the function given by Eq. (\ref{eq:25}). The averaging is
performed in the convential manner using the relations

$$e^{\hat{A}}e^{\hat{B}}=e^{\hat{A}+\hat{B}+(1/2)[\hat{A},\hat{B}]},\;\;\; \langle
e^{[F(b^+,b)]}\rangle=e^{(1/2)\langle F^2(b^+,b)\rangle},$$ where $F$ is an arbitrary
linear combination of boson operators. As a result, we find that,  at large times
$\varepsilon_Ft\gg 1$, the function in Eq.(\ref{eq:8.11}) $$\mathcal{G}_d(t)\sim
\mathcal{G}^0_d(t)t^{-\alpha_d}.$$ Consequently, we obtain in the energy
representation the expression
\begin{equation}
\mathcal{G}_d(\varepsilon)=Ae^{-i\pi\alpha_S}\Gamma(1-\alpha_S)\left[\frac{\hat{\tau}_0-\hat{\tau}_x}
{\varepsilon-\Delta+i\Gamma_{K}}\left(\frac{\varepsilon-\Delta+i\Gamma_{K}}{\varepsilon_0}\right)^{\alpha_S}+
\frac{\hat{\tau}_0+\hat{\tau}_x}
{\varepsilon-\Delta}\left(\frac{\varepsilon-\Delta}{\varepsilon_0}\right)^{\alpha_S}\right];\;\;\;
\varepsilon-\Delta>0, \label{eq:8.12}
\end{equation}
Here, $A\sim \varepsilon_F^{-1}$, $\Gamma(x)$ is the gamma function,
$\alpha_S=(\delta_{S}/\pi)^2$ и $\delta_{S}\sim \tilde{J}^2_z$ is the phase shift due
to the screening interaction in the S-channel, $\Delta=-\varepsilon_J$. The cut-off
parameter $ \varepsilon_0\sim \varepsilon_F$ since the velocity of excitations in the
collective channels is $v_F$, as follows from the Hamiltonian (\ref{eq:15}).

It should be noticed that the power-law dependence (\ref{eq:8.12}) occurs when the
condition $T_K\gg |\varepsilon_J|$ is failed. In this condition the Kondo temperature
is defined on the Emery-Kivelson line, $T_K\sim
\varepsilon_F\exp[-1/(J_{\bot}\rho_{0})]$. Using the expressions for
$T_K,|\varepsilon_J|$, we find that the power-law dependence (\ref{eq:8.12}) takes
place under condition $(J_{\bot}/\varepsilon_F)\ll
1/\ln(\varepsilon_F/|\tilde{J}_z|)$, $\tilde{J}_z\equiv (1/2\pi)(J_z-8v_F)$. When the
inverse inequality is satisfied, one may use the expression (\ref{eq:25}) on the
Emery-Kivelson line.

\section{Calculation of the density of states and self-energy functions}

{\bf 1.} The Green functions of impurity degrees of freedom for low energies are
defined by relations (\ref{eq:25}) or (\ref{eq:8.12}). Thus, to determine the
many-particle density of states for low energies in the presence of a pseudospin
impurity, one needs to substitute these Green functions into (\ref{eq:23}).

First, we must know the quantity
$\sum_k|V^{\alpha}_{kd}|^2\mathcal{G}^2_{\alpha\sigma}(k;\varepsilon)\approx
|V^{\alpha}_{k_F d}|^2\sum_k\mathcal{G}^2_{\alpha\sigma}(k;\varepsilon)$, where
$\sum_k\mathcal{G}^2_{\alpha\sigma}(k;\varepsilon)=-\partial
\Sigma_{0\alpha\sigma}(\varepsilon)/\partial\varepsilon$. Functions
$\Sigma_{0\alpha\sigma}(\varepsilon)$ are defined by expressions (\ref{eq:19}) with
the density of states
$\rho_{\alpha\sigma}(\varepsilon)=\rho_{0\alpha\sigma}(\varepsilon)$ from
(\ref{eq:28}); namely,
\begin{equation}
\Sigma^{(+)}_{0\alpha\sigma}(z)=\int\limits^{+\varepsilon_0}_{-\varepsilon_0}
d\varepsilon'
\frac{\rho_{0\alpha\sigma}(\varepsilon')}{z-\varepsilon'}=
\frac{1}{\Gamma(1+\eta_c)\varepsilon^{1+\eta_c}_0}\Big[\int\limits_0^{\varepsilon_0}
d\varepsilon'
\frac{(\varepsilon')^{\eta_c}}{z_{+}-\varepsilon'}+\int\limits_0^{\varepsilon_0}
d\varepsilon'
\frac{(\varepsilon')^{\eta_c}}{z_{-}+\varepsilon'}\Big]=A_0
(J_{0-}+J_{0+}); \label{eq:27a}
\end{equation}
Here, $z_{\pm}\equiv \varepsilon\pm i\gamma$, $\varepsilon>0$. For $\varepsilon<0$
имеем $\Sigma^{(-)}_{0\alpha\sigma}(z)=-\Sigma^{(+)}_{0\alpha\sigma}(z)$. Integrals
$J_{0\mp}(z)$, which are Gilbert transforms of the density of states for a Luttinger
liquid, are defined as $$J_{0\mp}(z)=\frac{\varepsilon_0^{\eta_c+1}}{(\eta_c+1)z}
F(1,\eta_c+1;\eta_c+2;\pm\frac{\varepsilon_0}{z_{\pm}}),$$ where
$F(\alpha,\beta;\gamma;x)$ is a hypergeometric function. Since we consider here only
the case when $\varepsilon_0 /|z|\gg 1$, we can use the transformation formulas for a
hypergeometric function with $|x|\gg 1$  \cite{GR} and obtain the following
expressions:
$$F(\alpha,\beta;\gamma;x)=\frac{\Gamma(\gamma)\Gamma(\beta-\alpha)}{\Gamma(\beta)\Gamma(\gamma-\alpha)}
\cdot
(-1)^{\alpha}x^{-\alpha}+\frac{\Gamma(\gamma)\Gamma(-\beta+\alpha)}{\Gamma(\alpha)\Gamma(\gamma-\beta)}
\cdot (-1)^{\beta}x^{-\beta},$$ We have used the familiar relations
$\Gamma(1)=1;\;\;\Gamma(\eta_c+1)=\eta_c\Gamma(\eta_c);\;\;
\Gamma(1-\eta_c)\Gamma(\eta_c)=\pi/\sin(\pi\eta_c)$. Thus, in long-wave limit, for
$\gamma\rightarrow 0$, we obtain
\begin{equation}
\Sigma^{(+)}_{0\alpha\sigma}(\varepsilon)=
\Big(\frac{2i\pi\sin[(\pi/2)\eta_c]\exp[i(\pi/2)\eta_c]}{\sin(\pi\eta_c)\Gamma(1+\eta_c)\varepsilon^{1+\eta_c}_0}\Big)
\cdot \varepsilon^{\eta_c};\label{eq:27aa}
\end{equation}
Since $F(1,1;2;\mp x)=\mp x^{-1}\ln(1\pm x)$, expression
(\ref{eq:27aa}) for $\eta_c=0$ corresponds to the principal term
in the expansion in parameter $x=(\varepsilon_0/\varepsilon)\gg
1$. Using the expressions
$$\sum_k\mathcal{G}^2_{\alpha\sigma}(k;\varepsilon)=-\frac{\partial
\Sigma^{(+)}_{0\alpha\sigma}(\varepsilon)}{\partial\varepsilon},\;\;\varepsilon>0;\;\;\;\;\;
\sum_k\mathcal{G}^2_{\alpha\sigma}(k;\varepsilon)=-\frac{\partial
\Sigma^{(-)}_{0\alpha\sigma}(\varepsilon)}{\partial\varepsilon}=-\frac{\partial
\Sigma^{(+)}_{0\alpha\sigma}(\varepsilon)}{\partial|\varepsilon|},\;\;\varepsilon<0,$$
we obtain with the help of formulas(\ref{eq:27a}) and
(\ref{eq:27aa})
\begin{equation}
\sum_k
\mathcal{G}^2_{\alpha\sigma}(k;\varepsilon)=\sum_k\mathcal{G}^2_{\alpha\sigma}(k;|\varepsilon|)=
\Big(\frac{-2i\exp[i(\pi/2)\eta_c]}{\Gamma(1+\eta_c)\varepsilon^{1+\eta_c}_0}\Big)\Big(
\frac{\pi\eta_c\sin[(\pi/2)\eta_c]}{\sin(\pi\eta_c)}\Big)\cdot
|\varepsilon|^{\eta_c-1}; \label{eq:29}
\end{equation}

For the Green function (\ref{eq:25}) (or, what is the same, on the
Emery-Kivelson line), substituting the expression (\ref{eq:29})
into (\ref{eq:23}), we arrive at the following expression for the
impurity contribution to the density of states:
\begin{equation}
\rho_{d\alpha}(|\varepsilon|)=\Big[\frac{2V^2\eta_c\sin[(\pi/2)\eta_c]}{\Gamma(1+\eta_c)
\varepsilon^{1+\eta_c}_0\sin(\pi\eta_c)}\Big]\Big[
\frac{\cos[(\pi/2)\eta_c]}{\Gamma^2_K}\cdot
|\varepsilon|^{\eta_c}+\frac{\sin[(\pi/2)\eta_c]}{\Gamma_K}\cdot
|\varepsilon|^{\eta_c-1}+\cos[(\pi/2)\eta_c]\cdot
|\varepsilon|^{\eta_c-2}\Big],
\label{eq:30}
\end{equation}

Here, we have introduced the notation $V^2\equiv
|V_{dk_F}^{\alpha}|^2$. For low energies, the most singular term
in the density of states is $\rho_{d\alpha}(|\varepsilon|)\sim
|\varepsilon|^{\eta_c-2}$. Pay attention to the fact that this
contribution is due to the term of the form ${\rm Re}
\mathcal{G}_d(\varepsilon)\cdot {\rm
Im}\sum_k\mathcal{G}^2_{\alpha\sigma}(k;\varepsilon)$ in ${\rm Re}
\mathcal{G}_d(\varepsilon)$ we take the part corresponding to
nonhybridized degrees of freedom (second term in relation
(\ref{eq:25})). While deriving formula (\ref{eq:30}), we took into
account the fact that the term with the $\delta$-function in ${\rm
Im} \mathcal{G}_d(\varepsilon)$ makes zero contribution to the
density of states for $\eta_c\neq 0$. Thus, we see that, in
contrast to noninteracting Fermi gas, excitations in a Luttinger
liquid considerably modify the peaks at the Fermi level, which are
generated by the interaction with a two-level impurity. The
singularity in the impurity density of states is enhanced for
$\eta_c<1$; on the contrary, it is suppressed for $\eta_c>2$.

{\bf 2.} The contribution to $\Sigma_{\alpha}(z)$ from the impurity term
$\rho_{d\alpha}(\varepsilon)$ to the density of states is determined by the expression

\begin{equation}
\Sigma_{\alpha}(z)=\int\limits^{+\infty}_{-\infty} d\varepsilon
\frac{\rho_{d\alpha}(\varepsilon)}{z-\varepsilon}=
\int\limits_0^{\infty} d\varepsilon
\frac{\rho_{d\alpha}(|\varepsilon|)}{z_{+}-\varepsilon}+\int\limits_0^{\infty}
d\varepsilon
\frac{\rho_{d\alpha}(|\varepsilon|)}{z_{-}+\varepsilon}\equiv
\Sigma^{(+)}_{\alpha}(z_+)+\Sigma^{(-)}_{\alpha}(z_-).
\label{eq:31}
\end{equation}

Taking into account expression (\ref{eq:30}) for the impurity
density of states, we see that $\Sigma_{\alpha}(z)$  contains the
integrals of three types, which are defined in the complex plane:
$$I_1^{(\mp)}=\int\limits^{+\infty}_{0} d\varepsilon
\frac{\varepsilon^{\eta_c-2}}{z\mp\varepsilon};\;\;\;I_2^{(\mp)}=\int\limits^{+\infty}_{0}
d\varepsilon
\frac{\varepsilon^{\eta_c-1}}{z\mp\varepsilon};\;\;\;I_3^{(\mp)}=\int\limits^{\varepsilon_0}_{0}
d\varepsilon \frac{\varepsilon^{\eta_c}}{z\mp\varepsilon}.$$

Evaluating integrals $I_1^{(\mp)}$, $I_2^{(\mp)}$ (which are
singular for small z) and substituting the obtained expressions
into (\ref{eq:31}), we obtain
\begin{equation}
\begin{split}
&(1)\;\;\; \text{for}\;\;\; \rho_{d\alpha}\sim
|\varepsilon|^{\eta_c-2};\;\;\;\Sigma_{\alpha}^{(\pm)}=A_1\cdot
z^{\eta_c-2}_{\pm}\cdot P_{\pm};\\
&P_{\pm}=-e^{-i\pi\eta_c},(-1)\;\;\text{for}\;\;
"\pm";\;\;\;A_1=\frac{2\pi\eta_c\cos(\pi\eta_c/2)\sin(\pi\eta_c/2)V^2}{\varepsilon^{\eta_c+1}_0\sin(\pi\eta_c)\Gamma(\eta_c+1)};
\end{split}
\label{eq:32a}
\end{equation}
\begin{equation}
\begin{split}
&(2)\;\;\; \text{for}\;\;\; \rho_{d\alpha}\sim
|\varepsilon|^{\eta_c-1};\;\;\;\Sigma_{\alpha}^{(\pm)}=A_2\cdot
z^{\eta_c-1}_{\pm}\cdot P_{\pm};\\
&P_{\pm}=-e^{-i\pi\eta_c},(+1)\;\;\text{for}\;\;
"\pm";\;\;\;A_2=\frac{2\pi\eta_c\sin^2(\pi\eta_c/2)V^2}{\Gamma_K\sin^2(\pi\eta_c)\Gamma(\eta_c+1)\varepsilon^{\eta_c+1}_0};
\end{split}
\label{eq:32b}
\end{equation}

Integrals $I_3^{(\mp)}(z)$ are defined by the same formulas that
were derived while determining $\Sigma_{0\alpha}^{(\pm)}$.

{\bf 3.} Beyond the Emery-Kivelson line, expression (\ref{eq:8.12}) defines the
retarded Green function $\mathcal{G}^R_{d}(\tilde{\varepsilon})$,
$\tilde{\varepsilon}=\varepsilon-\Delta$.  Let us consider the case when
$\Big[\sum_k\mathcal{G}^R_{\alpha\sigma}(k;\varepsilon)\Big]^2\sim
\varepsilon^{\eta_c-1}$, as a function of energy varies much more slowly than
$\mathcal{G}^R_{d}(\tilde{\varepsilon})$. The definitions of these two quantities
imply that this is possible for $\eta_c\gg \alpha_S$. In this case, we can set
$\Big[\sum_k\mathcal{G}^R_{\alpha\sigma}(k;\varepsilon)\Big]^2\approx
\Big[\sum_k\mathcal{G}^R_{\alpha\sigma}(k;\Delta)\Big]^2$.

For the nonhybridized degrees of freedom (second term in the expression
(\ref{eq:8.12})), it can easily be verified that the density of states for has the
form

\begin{equation} \label{eq:33}
\rho_d(\tilde{\varepsilon})=Q_d\Big(\frac{2V^2}{\pi\varepsilon_0^3}\Big)
\Big(\frac{\varepsilon_0}{\tilde{\varepsilon}}\Big)^{1-\alpha_S};\;\;\;\tilde{\varepsilon}>0;\;\;\;
Q_d=\Big(\frac{\pi^2}{\Gamma(\alpha_d)\Gamma(\eta_c)}\Big)
\Big(\frac{\varepsilon_0}{\tilde{\Delta}}\Big)^{1-\eta_c}.
\end{equation}
The retarded self-energy functions are defined as
\begin{equation}
\Sigma^R_{\alpha}(z)=\int\limits^{+\infty}_{-\infty} d\varepsilon
\frac{\rho_d(\varepsilon)}{z_+-\varepsilon}=\int\limits_{\tilde{\Delta}}^{\infty}
d\varepsilon \frac{\rho_d(\varepsilon)}{z_+-\varepsilon}+
\int\limits_{-\infty}^{\tilde{\Delta}} d\varepsilon
\frac{\rho_d(\varepsilon)}{z_+-\varepsilon}=\int\limits_0^{\infty}
d\tilde{\varepsilon}
\Big[\frac{\rho_d(\tilde{\varepsilon})}{\tilde{z}_+-\tilde{\varepsilon}}+
\frac{\rho_d(-\tilde{\varepsilon})}{\tilde{z}_+
+\tilde{\varepsilon}}\Big],\label{eq:33a}
\end{equation}
where $\tilde{z}_+=z_+ -\Delta;\;\;\;{\rm Re}\tilde{z}_+=\tilde{\varepsilon}>0$. Using
formulas (\ref{eq:33a}) and the expressions for the integrals derived above, we obtain
\begin{equation}
\Sigma^R_{\alpha}(z)=-\tilde{Q}_d\Big(\frac{4V^2}{\varepsilon_0^3}\Big)\cdot
\Big(\frac{\varepsilon_0}{\tilde{z}}\Big)^{1-\alpha_S};\;\;\;\tilde{Q}_d=Q_d
\ctg(\pi\alpha_S). \label{eq:33b}
\end{equation}
Since $\alpha_S<1$, the above contribution of the impurity density of states to
$\Sigma^R_{\alpha}(z)$ is also singular for small $|z|$ .

\section {FERMI-LIQUID RESONANCES IN THE VICINITY OF THE FERMI
LEVEL}

Let us now consider the Fermi-liquid singularities in the total
Green function (\ref{eq:19}), which are associated with the
scattering of many-particle excitations, described by Hamiltonian
Hsc in expression (\ref{eq:10a}). In the presence of resonance and
potential scattering, the position of the poles is determined from
the solution of Eq. (\ref{eq:20}). Substituting $\Sigma_{sc}$ into
Eq. (\ref{eq:20}) and assuming that all matrix elements appearing
in $\Sigma^{ab}_{\alpha}$ are determined by their values for $k =
k_F$, we obtain the equation
\begin{equation} \label{eq:34}
1-\mathcal{T}^2 \Sigma_L(z_r) \Sigma_R(z_r)-V^2
\mathcal{T}\mathcal{G}_d(z_r)\Sigma_L(z_r)
\Sigma_R(z_r)\Big[2+\mathcal{T}\Big(\Sigma_L(z_r)+\Sigma_R(z_r)\Big)\Big]=0,
\end{equation}
where $V$ and $\mathcal{T}$ are the matrix elements of the resonance and potential
scattering and $\Sigma_{\alpha}$ are the self-energy functions calculated in
(\ref{eq:32a}), (\ref{eq:32b}), and (\ref{eq:33b}). The three terms on the left-hand
side of Eq. (\ref{eq:34}), containing self-energy functions $\Sigma_{\alpha}$,
describe different scattering processes of many-particle excitations. The term
proportional to $\mathcal{T}^2$ corresponds to scattering of right fermions from left
fermions taking into account the fact that the density of states contains the impurity
contribution $\rho_{d\alpha}$. The terms proportional to $V^2$ and $\mathcal{T}^ 2$,
$V^2$ describe two possible processes of resonant scattering, i.e., scattering of
charge and pseudospin densities from many-particle impurity degrees of freedom, which
are described by the Green functions $\mathcal{G}_d(z_r)$ defined in (\ref{eq:25}),
(\ref{eq:8.12}).

{\bf 1.} On Emery-Kivelson line, the impurity density of states is defined by
expression (\ref{eq:30}); accordingly, the poles in the vicinity of the Fermi level
are generated by the contributions to $\Sigma_{\alpha}$, which are most singular in a
certain range of parameters for small $|z|\ll \Gamma_K$. We will first consider the
solutions to Eq. (\ref{eq:34}), which are associated with scattering of nonhybridized
excitations with band collective excitations of impurity degrees of freedom. In
accordance with expression (\ref{eq:25}), nonhybridized degrees of freedom are
described by the Green function $\mathcal{G}_d(z_r)=1/z_r$. We begin with the case
when the main contribution to $\Sigma_{\alpha}(z)$  comes from the term $\sim
|\varepsilon|^{\eta_c-2}$ in the density of states. Retaining in Eq. (\ref{eq:34}) the
principal terms with small values of $|z|$ , we obtain the following equation for the
poles:
\begin{equation} \label{eq:35}
1-V^2 \mathcal{T}^2\mathcal{G}_d(z_r)\Sigma_L(z_r)
\Sigma_R(z_r)\Big(\Sigma_L(z_r)+\Sigma_R(z_r)\Big)=0,
\end{equation}
For $z_{r\pm}$, we substitute $\Sigma^{\pm}_{\alpha}(z_{r\pm})$, respectively, into
Eq. (\ref{eq:35}). It should also be noted that $\Sigma_L(z_r)=\Sigma_R(z_r)$ since
$\rho_{L}(|\varepsilon|)=\rho_{R}(|\varepsilon|)$. Let us consider solutions with
$z_{r-}$. We write $z_{r-}$ in the form $z_{r-} = |z_{r-}|exp(i\varphi)$, physical
solutions corresponding to values of $0 <\varphi<\pi/2$. Solving the imaginary and
real parts of Eq. (\ref{eq:35}), we obtain the following solution with $|z_{r–}|\ll
\Gamma_K$:
\begin{equation}
\frac{|z_{r-}|}{\varepsilon_0}\approx A_1
\Big(\frac{\mathcal{T}}{\varepsilon_0}\Big)^{2/(7-3\eta_c)}\!\!\!\!\cdot
\Big(\frac{V}{\varepsilon_0}\Big)^{8/(7-3\eta_c)},\;\;\;\varphi=\frac{\pi}{7-3\eta_c},
\label{eq:36}
\end{equation}
where $A_1$ is a factor on the order of unity. This solution
exists when the following conditions are satisfied:
\begin{equation}
1<\eta_c<\frac{5}{3};\;\;\;\;
\Big(\frac{\mathcal{T}}{\varepsilon_0}\Big)^{1/4}\!\!\!\!\cdot
\Big(\frac{V}{\varepsilon_0}\Big)\ll\Big(\frac{\Gamma_K}{\varepsilon_0}\Big)^{(7-3\eta_c)/8}.
\label{eq:37}
\end{equation}

The first system of inequalities emerges from the requirement of
vanishing of the imaginary part of Eq. (\ref{eq:35}). When
conditions (\ref{eq:37}) are violated, Eq. (\ref{eq:35}) may have
solutions due to two other contributions to the density of states.
For $\rho_{d\alpha}\sim |\varepsilon|^{\eta_c-1}$ and
$\mathcal{G}_d(z_r)=1/z_r$ , there exists a solution
$z_{r-}=\varepsilon_{r-}<0$, corresponding to a localized level
below the Fermi energy. The position of this level is determined
by the energy
\begin{equation}
\frac{|\varepsilon_r|}{\varepsilon_0}=A_2\Big(\frac{V}{\varepsilon_0}\Big)^{8/(4-3\eta_c)}\!\!\!
\cdot\Big(\frac{\mathcal{T}}{\varepsilon_0}\Big)^{2/(4-3\eta_c)}\!\!\!\!\cdot
\Big(\frac{\varepsilon_0}{\Gamma_K}\Big)^{3/(4-3\eta_c)};\;\;\;|\varepsilon_r|\ll
\Gamma_K. \label{eq:38}
\end{equation}

Finally, for $\rho_{d\alpha}\sim |\varepsilon|^{\eta_c}$ and
$\mathcal{G}_d(z_r)=1/z_r$, we have a resonance with $z_{r+}$,
\begin{equation}
\frac{|z_{r+}|}{\varepsilon_0}\approx A_3
\Big(\frac{\mathcal{T}}{\varepsilon_0}\Big)^{1/(1-2\eta_c)}\!\!\!\!\cdot
\Big(\frac{V}{\varepsilon_0}\Big)^{6/(1-2\eta_c)}\!\!\!\!\cdot
\Big(\frac{\varepsilon_0}{\Gamma_K}\Big)^{4/(1-2\eta_c)},\;\;\;\varphi=\frac{2\pi\eta_c}{1-2\eta_c},
\label{eq:39}
\end{equation}
which exists when the following conditions are satisfied:
\begin{equation}
\eta_c<\frac{1}{6};\;\;\;\;
\Big(\frac{\mathcal{T}}{\varepsilon_0}\Big)^{1/4}\!\!\!\!\cdot
\Big(\frac{V}{\varepsilon_0}\Big)\ll\Big(\frac{\Gamma_K}{\varepsilon_0}\Big)^{(5-2\eta_c)/8}.
\label{eq:40}
\end{equation}
It can be seen that all the resonance obtained above are formed as
a result of scattering of nonhybridized impurity degrees of
freedom by many-particle excitations, which form the density of
states at low energies. The scattering of many-particle
excitations with $\mathcal{G}_d(z_r)=1/(z_{r\pm}+i\Gamma_K
\text{sign}\;\!\varepsilon_{r\pm})$, which is described by the
term in $\Sigma_{sc}$ proportional to $\mathcal{T}^2 V^2$, does
not lead to the formation of resonances with $|z_{r}|\ll
\Gamma_K$. In other words, Eq. (\ref{eq:35}) has no physical
solutions in this case.

Beyond the Emery-Kivelson line expressions for $\mathcal{G}_d$ and $\Sigma_{\alpha}$
from (\ref{eq:8.12}) and (\ref{eq:33b}) should be substituted into Eq. (\ref{eq:35}).
At ${\rm Re}(\tilde{z}_r)>0$ the position and width of new resonance are determined by
the formulas (we consider only contribution of the nohybridized impurity degrees of
freedom, which is described by the second term in Eq. (\ref{eq:8.12}))

\begin{equation}
\frac{|\tilde{z}_r|}{\varepsilon_0}\approx A_3
\Big(\frac{\mathcal{T}}{\varepsilon_0}\Big)^{1/2(1-\alpha_S)}\cdot
\Big(\frac{V}{\varepsilon_0}\Big)^{2/(1-\alpha_S)}\cdot
\Big(\frac{\Delta}{\varepsilon_0}\Big)^{(\eta_c-1)/4(1-\alpha_S)},\;\;\;\varphi=\frac{\pi}{4},
\label{eq:41}
\end{equation}
here $A_3\sim 1$. Since $\Delta\ll \varepsilon_0$ and $\eta_c\gg \alpha_S$, the
resonance for $\eta_c>1$ can be quite narrow with a width $\ll
\lambda_z/\varepsilon_F$.

{\bf 2.} In the range of parameters where Eq. (\ref{eq:35}) has no
solutions, the poles can be due to other scattering processes
apart from those making contribution to $\Sigma_{sc}$,
proportional $\mathcal{T}^2 V^2$. Assuming that this contribution
is small in the regions where Eq. (\ref{eq:35}) has no solutions,
we consider the equation
\begin{equation} \label{eq:42}
1-2V^2 \mathcal{T}\mathcal{G}_d(z_r)\Sigma_L(z_r) \Sigma_R(z_r)=0
\end{equation}

On Emery-Kivelson line, when the main contribution to $\Sigma_{\alpha}$ comes from the
term proportional to $\sim |\varepsilon|^{\eta_c-2}$ in the density of states, the
Fermi-liquid resonance is formed due to scattering of many-particle excitations by the
resonant level with $$\mathcal{G}_d(z_r)=\frac{1}{(z_{r\pm}+i\Gamma_K
\text{sign}\;\!\varepsilon_{r\pm})}\approx \mp \frac{i}{\Gamma_K},\;\text{for}\;{\rm
Re}(z_r)^{>}\!\!\!_{<}0.$$ The position and width of the Fermi-liquid resonance are
determined by the solution to Eq. (\ref{eq:42}),
\begin{equation}
\frac{|z_{r-}|}{\varepsilon_0}\approx A
\Big(\frac{\mathcal{T}}{\varepsilon_0}\Big)^{1/2(2-\eta_c)}\!\!\!\!\cdot
\Big(\frac{V}{\varepsilon_0}\Big)^{6/2(2-\eta_c)}\!\!\!\!\cdot
\Big(\frac{\varepsilon_0}{\Gamma_K}\Big)^{1/2(2-\eta_c)},\;\;\;\varphi=\frac{\pi}{4(2-\eta_c)},
\label{eq:43}
\end{equation}
where $A$ is a factor on the order of unity. This solution exists
when the following conditions are satisfied:
\begin{equation}
1<\eta_c<\frac{3}{2};\;\;\;\;
\Big(\frac{\mathcal{T}}{\varepsilon_0}\Big)^{1/6}\!\!\!\!\cdot
\Big(\frac{V}{\varepsilon_0}\Big)\ll\Big(\frac{\Gamma_K}{\varepsilon_0}\Big)^{(5-2\eta_c)/6}.
\label{eq:44}
\end{equation}
The same scattering process generates the Fermi-liquid resonance
with $z_{r–}$ in the case when the main contribution to
$\Sigma_{\alpha}$ comes from the term proportional to $\sim
|\varepsilon|^{\eta_c-1}$ in the density of states. In the present
case, the new resonance exists for
\begin{equation}
\eta_c<1;\;\;\;\;\;
\Big(\frac{\mathcal{T}}{\varepsilon_0}\Big)^{1/6}\!\!\!\!\cdot
\Big(\frac{V}{\varepsilon_0}\Big)\ll\Big(\frac{\Gamma_K}{\varepsilon_0}\Big)^{(3-2\eta_c)/6}.
\label{eq:44a}
\end{equation}

Finally, new singularities may appear due to potential scattering
of right fermions from left ones taking into account the impurity
density of states (\ref{eq:30}). The positions of the poles in
this case is determined by the solutions to the equation
$1-\mathcal{T}^2\Sigma_L(z_r) \Sigma_R(z_r)=0$. It can easily be
verified that a solution exists in the case when the main
contribution to  $\Sigma_{\alpha}$  comes from the term
proportional to $\sim |\varepsilon|^{\eta_c-2}$ and has the form
of a localized level below the Fermi energy.

It can also be proved that potential scattering beyond the Emery-Kivelson line
generates a localized level above the Fermi energy.

If the density of states at the Fermi level is determined by
expression $\rho_{0\alpha}$ for a pure Luttinger liquid, the
potential scattering of right fermions from left ones does not
lead to the formation of additional Fermi-liquid resonances or
levels at low energies. This case was treated in [10].

Concluding the section, let us prove that the inequalities for
parameter  $\eta_c$ derived above, which determine the ranges for
new Fermi-liquid resonances, are in accordance with the inequality
$K_c < 1/2$. It should be recalled that this is the condition of
applicability of the model with $J_B=0$. In particular, the
conditions for $\eta_c$ in (\ref{eq:37}) and (\ref{eq:44})
correspond to the inequalities $0.13 < K_c < 0.2$ and $0.11 < K_c
< 0.2$, respectively. Finally, the condition $\eta_c<1$  in
(\ref{eq:44a}) corresponds to values of $K_c < 0.2$. In this case,
the mechanism considered in this study ensures instability in the
region $1>\eta_c>1/8$, which is also in agreement with the
conditions derived above.

\section {Concluding remarks. The behavior of physical quantities}

Several types of NFL excitations with different quantum numbers characterize a
Luttinger liquid with two-level impurities at low energies. The results describe above
indicate that the scattering of many-particle excitations of various types from one
another leads to the emergence of an additional Fermi-liquid resonances in the
vinicity of the Fermi level in a 1D system. The conditions are dtermined, under which
the fermi-liquid states with a new energy scale muchsmaller than the Kondo temperature
is the ground state of the system.

The type of many-particle excitations and, accordingly, the type of the phase state
are determine by the following parameters: the value of anomalous dimensionality
$\eta_c$ in the charge channel of a Luttinger liquid (or, what is the same, the value
of parameter $K_c$) as well as the initial energy $E_d$ of a deep level. This energy
appears in exchange constants. Reducing (increasing) the depth of the impurity level
leads to increasing (reducing) the Kondo temperature. Let us define the phase states
of the system in dependence on the values of these parameters.

In accordance of conditions (\ref{eq:37}), (\ref{eq:40}), (\ref{eq:44}), and
(\ref{eq:44a}), {\bf new Fermi-liquid resonances are absent} when, at any rate,
$\eta_c$ is greater than $5/3$ and/or the Kondo temperature on the Emery-Kivelson line
is found to be very low. In the latter case, the impurity level is very deep and
conditions of the type $|z_r|\ll T_K$ are violated and we obtain the phase state of
the system , in which excitation of the Luttinger liquid take place in th charge
channel and the excitations generated by two-channel exchange interaction occur in the
pseudospin channel.

The low-temperature behavior of the heat capacity $C(T)$ is determined by the product
$\gamma T$, where $\gamma\sim \ln(T_K/T)$ in the second order of perturbation theory
in $\tilde{J}_z/J_{\bot}$ \cite{EK}, \cite{SG}. The logarithmic dependence of $C/T$
occurs instead of the constant value of $\gamma_c\sim u_c^{-1}$ in a pure Luttinger
liquid. The homogeneous static susceptibility behaves analogously on the
Emery-Kivelson line. According to the conditions described in conclusion of section
IV, the logarithmic dependence is transformed into the power dependence $\gamma\sim
T^{-1+\alpha_S}$ occuring beyond the Emery-Kivelson line at the crossover temperature
$T_{c}\sim |\varepsilon_J|$. If the initial values of the parameters is such that the
condition $T_K\gg |\varepsilon_J|$ holds, the crossover takes place at deepening the
impurity level.

In the phase state considered here, anomalous correlations exist at the impurity
site.The divergence of the correlator $<S^+S^->(\omega=0,T)\sim \ln(1/T)$ corresponds
to free rotation of th impurity pseudospin. We recall here that in an atomic Fermi gas
pseudospin fluctuations describe the fluctuations of the relative density at two
levels corresponding to intrinsic states of atoms. The anomalous behavior of the
one-site correlator $<S^+S^->$ corresponds to anomalous increase (upon cooling) of the
correlations between the occupancies for two intrinsic states of an impurity site.
This made indicate a tendency to the formation of the superfluid stae in the relative
density of two component of the impurity subsystem in a quasi-one-dimensional system.

On the Emery-Kivelson line {\bf Fermi-liquid resonances exist} for relatively small
values of $\eta_c$ and high values of $T_K$ in accordance with the formulas derived
above. If the value of $\eta_c$ satisfiesany condition of the existence of resonances,
but the corresponding condition of the type $|z_r|\ll T_K$ is violated, we can obtain
an additional Fermi-liquid resonance by reducing the depth of the impurity level and,
hence, by increasing the Kondo temperature.

Expression (\ref{eq:41}) implies that {\bf Fermi-liquid resonances exist} beyond the
Emery-Kivelson line for all admissible values of $\alpha_S$ (i.e., impurity level
depths), but only for strong interactions in the charge channels of a Luttinger
liquid.

Both on and beyond the Emery-Kivelson line, transitions from an NFL state to an FL
state can take place. The characteristic crossover temperatures are $T_r\sim
\gamma_r$, where $\gamma_r$ are the widths of Fermi-liquid resonances.

It should be emphasized that the Fermi-liquid resonances lead to the existence of new
energy scales $\gamma_r$, which are much smaller than the Kondo temperature.

\section{Acknowledgment}

This study was supported by Russian Foundation for Basic Research.

\end{document}